\documentstyle[12pt]{article}

\begin{document}

\begin{titlepage}

\rightline{RU-92-45}
\rightline{BUHEP-92-37}
\rightline{November, 1992}

\begin{center}
{\LARGE\bf{A Gauge-fixed Hamiltonian for Lattice QCD}}

\bigskip
\bigskip
\bigskip

{\large J. B. Bronzan}

\smallskip

Department of Physics and Astronomy\\
Rutgers University \\
Piscataway, NJ 08855-0849 \\
email: bronzan@physics.rutgers.edu

\bigskip
\bigskip

{\large Timothy E. Vaughan}

\smallskip

Department of Physics \\
Boston University \\
590 Commonwealth Ave. \\
Boston, MA 02215 \\
email: tvaughan@buphyk.bu.edu

\end{center}

\bigskip

\abstract{
\noindent
We study the gauge fixing of lattice QCD in 2+1 dimensions, in the
Hamiltonian formulation.  The technique easily generalizes to other
theories and dimensions.  The Hamiltonian is rewritten in terms of
variables which are gauge invariant except under a single global
transformation.  This paper extends previous work, involving only pure
gauge theories, to include matter fields.}

\end{titlepage}

\section{Introduction}
\setcounter{equation}{0}
It is straightforward to write a Hamiltonian for lattice QCD, as was
originally derived by Kogut and Susskind in 1975.\cite{kogut75}
However, the Hamiltonian operator commutes with local gauge
transformations at each site on the lattice.  Since these operators
commute among themselves, the Hilbert space is split into subspaces
characterized by the color charge at each site.  In particular, the
vacuum subspace has zero color at every site.

Writing a general wave function belonging to a specific subspace is a
nontrivial task.  M\"uller and R\"uhl \cite{muller} showed how to do
this for the vacuum subspace of a pure gauge theory in $2+1$ dimensions,
and Bronzan \cite{bronzan85} extended their work to $3+1$ dimensions
and the two-charge subspace.  In this paper we will further extend this
work to include theories with matter fields.  We will illustrate our
technique using lattice QCD in $2+1$ dimensions, but it can be easily
generalized to other dimensions and other theories which have unitary
gauge groups.

The layout of the rest of the paper is as follows.  In Section~2 we will
review the approach of M\"uller and R\"uhl.  In Section~3 we will
present our method for gauge fixing in theories with fermions.  In
Section~4, we will summarize and comment on the generalization of the
technique to other theories.

\section{Gauge fixing pure SU(3)}
\setcounter{equation}{0}
We use a lattice consisting of $N\times N$ links, with open
boundary conditions.  We label links by the pair $({\bf s},\mu)$,
where ${\bf s}$ is the site from which the link starts, with $0\le
s_x,s_y\le N$, and $\mu$ is the direction of the link.  Each link is
associated with an element $U_{{\bf s},\mu}$ of the gauge group in the
usual fashion.  By ${\cal J}_{L_\alpha}({\bf s},\mu)$ we denote the
generators of the gauge group in differential form.  For the
eight-dimensional group SU(3), on which we will focus our attention,
$\alpha = 1,\ldots,8$.  The action of ${\cal J}_{L_\alpha}$ on a matrix
representation of SU(3) is to left-multiply it by the matrix form of the
$\alpha^{\rm th}$ generator.  For example,
\begin{equation}
{\cal J}_{L_\alpha}U_{\beta\gamma} = {1\over
2}(\lambda_{\alpha}U)_{\beta\gamma},
\end{equation}
where $\lambda_{\alpha}$ is the $\alpha^{\rm th}$ Gell-Mann matrix.
Similarly, we denote by ${\cal J}_{R_\alpha}({\bf s},\mu)$ the
right-multiplying generators.  The operators $-{\cal J}_L$ and ${\cal
J}_R$ obey the usual Lie algebra of SU(3).

The Hamiltonian can be written as the sum of an ``electric'' piece and
and a ``magnetic'' piece:
\begin{equation}
H = H_e + H_m,
\end{equation}
where
\begin{eqnarray}
H_e & = & {g^2\over{2a}}\sum_{{\bf s},\mu}\vec{\cal J}^2_L({\bf
s},\mu),\\
H_m & = & -{1\over{g^2a}}\sum_{{\bf s},\mu>\nu}[Tr(U_{{\bf s},\mu}U_{{\bf
s+\hat\mu},\nu}U^\dagger_{{\bf s+\hat\nu},\mu}U^\dagger_{{\bf s},\nu}) +
H.c. - 6].
\end{eqnarray}
The arrow symbolizes that $\vec{\cal J}_L$ is a vector on the SU(3)
manifold.

The gauge-fixing procedure of Ref.~\cite{muller} begins with a change of
variables.  First, define a ``maximal tree'' consisting of all $y$ links,
and all $x$ links on the $x$ axis.  (See Figure~1 of Ref.~\cite{muller}.)
All the links on the maximal tree remain in the new variable set.  The
other links of the lattice are replaced by ``loop variables'':
\begin{equation}
y_{\bf s} = \prod_{l\in R_{\bf s}}(U_l)^{\sigma_l},
\end{equation}
where we have used $l = ({\bf s},\mu)$ as shorthand.  Here, $R_{\bf s}$
represents an oriented path that follows
$(0,0)\rightarrow(0,s_y)\rightarrow(s_x,s_y)
\rightarrow(s_x,0)\rightarrow(0,0)$.  (See Figure~3 of Ref.~\cite{muller}.)
$\sigma_l = \pm 1$ depending on whether $R_{\bf s}$ passes through the
link in the positive or negative direction.  The choice of these
particular paths is discussed in Ref.~\cite{muller}.  Since SU(3) is a
non-Abelian group, it is important to note that the oriented path
gives the order in which the group elements are multiplied from left to
right.  While it is convenient, let us also define $V_{\bf s}$ as the
same ordered product, except that the path stops at ${\bf s}$
rather than returning to $(0,0)$.

Now, denote by ${\cal J}_{L_\alpha}(y_{\bf s})$ and ${\cal
J}_{R_\alpha}(y_{\bf s})$ the differential generators
for the corresponding loop variable.  Also, denote by
$D_{\beta\alpha}(V_{\bf s})$ the adjoint representation of SU(3) in the
basis where the rows and columns are labeled by the indices of the
generators.  We use Eq.~(4.13) of Ref.~\cite{bronzan85} to write
the old generators in terms of the new:
\begin{equation}
{\cal J}_{L_\alpha}(U_{{\bf s},\mu}) = \epsilon_{{\bf s},\mu}{\cal
J}_{L_\alpha}(U_{{\bf s},\mu}) + D_{\beta\alpha}(V_{\bf s})[\sum_{{\bf
s}'}^{\cal P}{\cal J}_{L_\beta}(y_{{\bf s}'}) - \sum_{{\bf s}'}^{\cal
N}{\cal J}_{R_\beta}(y_{{\bf s}'})], \label{eq:changevar}
\end{equation}
where the first sum is over loops which make transits of the
link $U_{{\bf s},\mu}$ in the positive direction, and the second sum is
over negative transits.  $\epsilon_{{\bf s},\mu} = 1$ if the link
is on the maximal tree, and 0 if it is not.  Note also that there is an
implicit sum over $\beta$.

With this expression in hand, we are now able to
write the electric Hamiltonian.
\begin{equation}
H_e = {{g^2}\over{2a}}\sum_{{\bf s},\mu}[\sum_{{\bf s}'}^{\cal P}\vec{\cal
J}_L(y_{{\bf s}'}) - \sum_{{\bf s}'}^{\cal N}\vec{\cal J}_R(y_{{\bf
s}'})]^2
\end{equation}
Note that we are
able to leave out the generators of the maximal tree variables, in
accordance with the discussion in Sec. V of Ref.~\cite{bronzan85}.

The magnetic Hamiltonian can also be written solely in terms of the
loop variables:
\begin{equation}
H_m =-{1\over{g^2a}}\sum_{s_x,s_y=1}^N
[Tr(y_{s_x-1,s_y}^{\dagger}y_{s_x-1,s_y-1}y_{s_x,s_y-1}^{\dagger}y_{s_xs_y})
+ H.c. - 6].
\end{equation}
In this expression, if either subscript of $y_{\bf s}$ is zero, then
$y_{\bf s}$ is defined to be the unit element.

This completes the gauge fixing of the SU(3) Hamiltonian.  The
general state in the zero-charge subspace is independent of the maximal
tree variables, and we have also eliminated these variables from the
Hamiltonian.  The sole vestige of local gauge invariance is the
invariance under a transformation performed at the origin.  Since this
transforms all the loop variables, it is in fact a global SU(3)
invariance.  The Hamiltonian is invariant under this transformation, and
physical wave functions in the subspace must also be invariant
under it.  Once we have imposed this condition, we have eliminated all
gauge arbitrariness from the problem.

\section{Gauge fixing QCD}
\setcounter{equation}{0}
In this section, we add quark fields to the Hamiltonian and carry out
the same prescription.  We will see that this requires an additional
change of variables to banish the unphysical degrees of freedom from the
Hamiltonian.

We add to the Hamiltonian a naive Dirac term:
\begin{equation}
H = H_e + H_m + H_q,
\end{equation}
\begin{equation}
H_q = {{a^3}\over 2}\sum_{{\bf s},\mu}[\overline\psi({\bf
s})\gamma_{\mu}U_{{\bf s},\mu}\psi({\bf s}+\hat\mu)
- \overline\psi({\bf s}+\hat\mu)\gamma_{\mu}U_{{\bf
s},\mu}^\dagger\psi({\bf s})] + ma^4\sum_{\bf s}\overline\psi({\bf
s})\psi({\bf s}).
\end{equation}
We need not be concerned here with the doubling problem.  The technique
will work if either Wilson or staggered fermions are implemented.

The initial variable set consists of the link variables $U_{{\bf
s},\mu}$ and the quark variables $\psi({\bf s})$ and $\psi^{\dagger}({\bf
s})$.  The first step is to change to a new set of quark variables,
\begin{eqnarray}
\psi'({\bf s}) = V_{\bf s}\psi({\bf s}), \nonumber \\
\psi'^{\dagger}({\bf s}) = \psi^{\dagger}({\bf s})V_{\bf s}^\dagger,
\label{eq:newquark}
\end{eqnarray}
where $V_{\bf s}$ is defined as it was in Section 2.  Note that since
$V_{\bf s}$ is unitary, $\psi'({\bf s})$ and $\psi'^{\dagger}({\bf
s})$are still canonical fermions.  The motivation behind this change of
variables, as we will see explicitly at the end of this section,  is
that the gauge element in $H_q$ will now be either a product of loop
variables, or the unit element.  Thus, we have banished the maximal tree
elements from these terms, as desired.

With this change of variables, the gauge generators and the quark
operators are no longer ``independent,'' by which we mean that in general
\begin{equation}
[{\cal J}_{L_\alpha}({\bf s}_1,\mu),\psi'({\bf s}_2)] \ne 0.
\end{equation}
We can eliminate this difficulty.  First, define the operator
\begin{equation}
{\cal F}_\alpha({\bf s}) = -{1\over 2}\psi'^{\dagger}({\bf
s})\lambda_\alpha\psi'({\bf s}),
\end{equation}
where $\lambda_\alpha$ is the $\alpha^{\rm th}$ Gell-Mann matrix.  Now,
define a new set of operators
\begin{equation}
{\cal J}_{L_\alpha}'({\bf s},\mu) = {\cal J}_{L_\alpha}({\bf
s},\mu) - D_{\beta\alpha}(V_{\bf s})\sum_{{\bf s}'}^{\cal P}{\cal
F}_\beta({\bf s}'),
\end{equation}
where we have indicated that the sum is over paths to ${\bf s}'$
which make a positive transit across $U_{{\bf s},\mu}$.  Recall that
$D_{\alpha\beta}(V_{\bf s})$ is the adjoint representation of SU(3).
The following properties of these new operators are easily verified:
\begin{enumerate}
\item ${\cal J}_L'$ obeys the same algebra as ${\cal J}_L$:
\begin{equation}
[{\cal J}_{L_\alpha}'({\bf s}_1,\mu_1),{\cal J}_{L_\beta}'({\bf
s}_2,\mu_2)] = -i\delta_{{\bf s}_1{\bf
s}_2}\delta_{\mu_1\mu_2}\sum_{\gamma}f_{\alpha\beta\gamma}{\cal
J}_{L_\gamma}'({\bf s}_1,\mu_1);
\end{equation}
\item ${\cal J}_L'$ commutes with the primed quark operators:
\begin{equation}
[{\cal J}_{L_\alpha}'({\bf s}_1,\mu),\psi'({\bf s}_2)] =
[{\cal J}_{L_\alpha}'({\bf s}_1,\mu),\psi'^{\dagger}({\bf s}_2)] = 0;
\end{equation}
\item the commutators of ${\cal J}_L'$ with an arbitrary function of group
elements is the same as that of ${\cal J}_L$:
\begin{equation}
[{\cal J}_{L_\alpha}'({\bf s},\mu),f(U_{{\bf s},\mu})] = [{\cal
J}_{L_\alpha}({\bf s},\mu),f(U_{{\bf s},\mu})].
\end{equation}
\end{enumerate}
This last point is important, because then the primed analogue of
Eq.~(\ref{eq:changevar}) will hold.

We now proceed with the second change of variables, which is
the link-to-loop change used in the pure gauge case.  As a result, the
operators in the electric Hamiltonian are
\begin{eqnarray}
{\cal J}_{L_\alpha}({\bf s},\mu) & = & \epsilon({\bf s},\mu){\cal
J}_{L_\alpha}'({\bf s},\mu) \nonumber \\
 & + & D_{\beta\alpha}(V_{\bf s})[\sum_{{\bf s}'}^{\cal P}{\cal
J}_{L_\beta}'(y_{{\bf s}'}) - \sum_{{\bf s}'}^{\cal N}{\cal
J}_{R_\beta}'(y_{{\bf s}'}) + \sum_{{\bf s}'}^{\cal P}{\cal
F}_\beta({\bf s}')].
\end{eqnarray}
We can thus write the electric Hamiltonian,
\begin{equation}
H_e = {{g^2}\over{2a}}\sum_{{\bf s},\mu}[\sum_{{\bf s}'}^{\cal
P}\vec{\cal J}_L'(y_{{\bf s}'}) - \sum_{{\bf s}'}^{\cal N}\vec{\cal
J}_R'(y_{{\bf s}'}) + \sum_{{\bf s}'}^{\cal P}\vec{\cal F}_{\beta}({\bf
s}')]^2.
\end{equation}
The magnetic Hamiltonian is unchanged from the pure-gauge case:
\begin{equation}
H_m =-{1\over{g^2a}}\sum_{s_x,s_y=1}^N
[Tr(y_{s_x-1,s_y}^{\dagger}y_{s_x-1,s_y-1}y_{s_x,s_y-1}^{\dagger}y_{s_xs_y})
+ H.c. - 6].
\end{equation}
The quark Hamiltonian is
\begin{eqnarray}
H_q & = & {{a^3}\over 2}\sum_{{\bf s},\mu}[\overline{\psi'}({\bf
s})\gamma_{\mu}V_{\bf s}^{\dagger}U_{{\bf s},\mu}V_{{\bf
s}+\hat\mu}\psi'({\bf s}+\hat\mu)
- \overline{\psi'}({\bf s}+\hat\mu)\gamma_{\mu}V_{{\bf
s}+\hat\mu}^{\dagger}U_{{\bf s},\mu}^{\dagger}V_{\bf s}\psi'({\bf s})]
\nonumber \\
 & & + ma^4\sum_{\bf s}\overline{\psi'}({\bf
s})\psi'({\bf s}).
\end{eqnarray}
When $\hat\mu = \hat x$, the gauge element in $H_q$ is
\begin{equation}
V_{\bf s}^{\dagger}U_{{\bf s},\mu}V_{{\bf s}+\hat\mu} = V_{{\bf
s}+\hat\mu}^{\dagger}V_{{\bf s}+\hat\mu} = 1;
\end{equation}
when $\hat\mu = \hat y$, the gauge element is
\begin{equation}
V_{\bf s}^{\dagger}U_{{\bf s},\mu}V_{{\bf s}+\hat\mu} = y_{\bf
s}y^\dagger_{{\bf s}+\hat\mu}.
\end{equation}
We see, then, that $H_q$ does not depend on maximal tree variables.
We have thus eliminated the maximal tree variables from all terms of the
Hamiltonian.  Again, the general wave function in the zero-charge
subspace can be written in terms of only the loop variables and the new
quark operators.  Here, too, we must impose global SU(3) invariance on
the wave function.  Once this is done, we have again eliminated all
gauge arbitrariness from the problem, as desired.

\section{Summary}
\setcounter{equation}{0}
We have presented a technique to fix the gauge of the lattice QCD
Hamiltonian in $2+1$ dimensions, in the zero-charge subspace.  As we
stated in the introduction, this method can be generalized.  The
generalization to $3+1$ dimensions and the two-charge subspace is
conceptually trivial, but does involve a more intricate definition of the
paths in the various changes of variables.  These paths are described in
Ref. \cite{bronzan85}.  One of us has used the scheme in $3+1$
dimensions in a calculation of the glueball spectrum. \cite{vaughan92a}

Generalization to other gauge groups is easily accomplished.  A
restriction is that the group must be unitary, so that the transformed
quarks defined in Eq.~(\ref{eq:newquark}) remain canonical fermions.

\bigskip

\noindent{\large{\bf Acknowledgements}}

This work was supported in part by the National Science Foundation under
grant number PHY 88-18535 and the U.~S. Department of Education.

\bibliographystyle{plain}

\end{document}